\documentstyle[12pt]{article}
\input epsf

\begin{document}

\topmargin 0pt
\oddsidemargin 5mm

\setcounter{page}{1}
\begin{titlepage}
\vspace{2cm}
\rightline{Preprint YerPhI-1495(12)-97}
\begin{center}

{\bf
 Higgs Bosons Production with Photons at Lepton-Antilepton  Colliders}\\
\vspace{5mm}
{\large R.A. Alanakyan}\\
\vspace{5mm}
{\em Theoretical Physics Department,
Yerevan Physics Institute,
Alikhanian Brothers St.2,

 Yerevan 375036, Armenia\\}
 {E-mail: Alanak @ lx2.YERPHI.AM\\}
\end{center}

\vspace{5mm}
\centerline{{\bf{Abstract}}}
By model independent way scalar and pseudoscalar neutral Higgs
boson production with photon
in  the tree process
$\mu^+\mu^- \rightarrow H^0 \gamma$  are considered.For the
Standard Model and Minimal Supersymmetric Standard Model cases
numerical estimates are obtained.The model independent flavour
changing Higgs bosons
production in the tree processes $e^+e^-,\mu^+e^- \rightarrow H^0_{fc} \gamma$
 is also considered.

\vspace{5mm}
\vfill
\centerline{{\bf{Yerevan Physics Institute}}}
\centerline{{\bf{Yerevan 1997}}}

\end{titlepage}

                 {\bf 1.Introduction}

 As known, Higgs bosons interactions with fermions are
 proportional to their masses.This property of Higgs bosons make
possible their production at $\mu^+\mu^-$-colliders
 in resonance \cite{C}, \cite{BBG} (for references on $\mu^+\mu^-$-colliders see
 ref.\cite{P} and references therein):
\begin{equation}
\label{A1}
 \mu^+\mu^- \rightarrow H^0.
 \end{equation}
Analogous process in $e^+e^-$-collisions is suppresed by the
smallness of electron mass.

However, mass of the Higgs bosons is the free
 parameter of the theory and, thus, we don't know  which energies are
necessary for  $H^0$-bosons production
 in resonance (i.e. at $s=m^2_H$).

Here we study scalar and pseudoscalar
Higgs boson production by model independent way
in  the processes
\footnote{Scalar and pseudoscalar Higgs bosons production in the
 process (2) has been considered previously in \cite{RA}
however the author of these references do not consider the effect of muon
mass (formula(6) below) and in numerical results for the total
cross section  cut angles near colliding beams direction, where
the cross section is divergent.}
\begin{equation}
\label{A2}
 \mu^+ \mu^-\rightarrow H^0\gamma,
		\end{equation}

as described by the diagrams of Fig.1,
where $H^0$-are scalar or pseudoscalar Higgs bosons (see Appendix
A for model independent interaction of Higgs bosons with fermions
as well as Higgs bosons interactions with fermions in the
framework of the Minimal Supersymmetric Standard Model(MSSM),
see \cite{HK,GH} and references therein).

Whereas at $e^+e^-$-colliders the tree contribution 
is supressed by the smallness of electron mass 
in comparision with loop contribution \cite{BPR},\cite{ABD},\cite{WY},at
$\mu^+ \mu^-$-colliders, as we will see below under some conditions the tree 
contriobution may exceed loop effects.

We also consider similar processes (see Fig.1)
\footnote{for references on $\mu^+e^-$-colliders see
 \cite{HW} and references therein}:
\begin{equation}
\label{A3}
 \mu^+ e^-\rightarrow H_{fc}^0\gamma,
\end{equation}

\begin{equation}
\label{A4}
 \mu^+\mu^-,e^+ e^-\rightarrow H_{fc}^0\gamma,
\end{equation}

 where $H^0_{fc}$-are flavour
 changing scalars or pseudoscalars \cite{FW}.
 Such particles are predicted
 by many extentions of the Standard Model, for example, they are
 contained into SUSY theories with  $R$-parity violation
 \cite{F}-\cite{D} (the role
 of $H^0_{fc}$- bosons in such theories plays scalar
 neutrino ).

 In Appendix A the
 most general interaction of $H^0_{fc}$-bosons with fermoions
 is presented.It must be noted that in some models (in particular
 in models with $R$-parity violation ) Yucava couplings are not
 necessarily proportional to the fermion masses and consequently
 the processes (3),(4) may  also be measurable as we see below at
 sufficiently large couplings.

Flavour changing Higgs bosons exchanges may
 contribute to the processes with lepton flavour violation.In
 \cite{CHWY} has
 been considered their contribution to the decay $\mu \rightarrow
 e\gamma$, in \cite{M,HM,HW},
 their contribution to the muonium-antimuonium conversion
 ( for details of muonium -antimuonium experiments see \cite{PSI}).

 Flavour changing scalars and pseudoscalars may also be produced
 virtually or in resonance in $\mu^+e^-$-collisions \cite{HW},
$\mu^+\mu^-$-collisions
,$e^+e^-$-collisions\cite{BGH}-\cite{KRSZ}.

 Although the cross section of the processes (2),(3) is
essentially smaller than the cross section of the resonant
process (1) and $ \mu^+e^- \rightarrow H^0_{fc}$,the
processes (2),(3) are allowed under more soft restriction
$\sqrt{s}>m_H$ where $m_H$ is the mass of the $H^0$- or
$H^0_{fc}$-bosons.

Produced in reactions (3),(4)  $H^0_{fc}$-bosons decay unambiguosly
into leptons ($H^0_{fc}\rightarrow\bar{l_i}l_j$) and consequently
there are no backgrounds to the process (3) with subsequent
decays of $H^0_{fc}$-bosons in contrast to the process  (2)
, where backgrounds like $ \mu^+ \mu^-\rightarrow \bar{b}b\gamma$
imitating appropriated Higgs boson decays
exist .The backgrounds are absent for the process (3) and resonant
production of $H^0_{fc}$-bosons  even if produced flavour changing scalar or
pseudoscalar decay into fermions of the same flavours.

{\bf 2. The $ \mu^+ \mu^-\rightarrow H^0\gamma$ process}

For the differential cross section for both scalar and
pseudoscalar Higgs boson cases we obtain the following result
\footnote{The cross sections for the scalar and pseudoscalar cases
are not equal to each other if we do not neglect terms
proportonal to $m^4_{\mu}$  and of higher order in $m_{\mu}$.}:

\begin{equation}
\label{A5}
\frac{d\sigma( \mu^+ \mu^-\rightarrow H^0\gamma)}{dt}=\frac{ \pi
\alpha^2}{4sin^2
 \theta_W} \frac{m^2_{\mu}}{m_W^2}F_{S,P}^2 \frac{1}{s^2}
(
\frac{(m^4_H+s^2)}{(t-m^2_{\mu})(u-m^2_{\mu})}-2m^2_{\mu}m^2_H(\frac{1}{(t-m^2_{\mu})^2}+\frac{1}{(u-m^2_{\mu})^2})).
\end{equation}

Here we use the following notations:
 $s=(k_1+k_2)^2$,$t=(k_1-k_3)^2$,$u=(k_2-k_3)^2$,$m_{\mu}$-mass of the muon.

If we neglect muon mass we obtain:
\begin{equation}
\label{A6}
\frac{d\sigma(\mu^+ \mu^-\rightarrow H^0\gamma) }{d\cos\theta
}=\frac{ \pi \alpha^2}{2\sin^2
\theta_W } \frac{m^2_{\mu}}{m_W^2}
F_{S,P}^2\frac{1}{s-m^2_H}\left(\frac{s^2+m_H^4}{s^2} \right) \frac{1}{\sin^2 \theta}.
\end{equation}
Here 
 $\theta $  is the angle between the photon momentum $\vec{k_3}$
 and  muon momentum $\vec{k_1}$.

As we see, our  result contains a collinear
singularity at $\theta=0,\pi$, however, photons with $\vec{k_3}$
nearly parallel to the beam direction can not be detected and we cut some
 cone  near this direction
as has been done
 for the  $e^+e^-\rightarrow Z^0\gamma$  process (see ref. \cite{B}  and
 references therein).

 It  must be noted however that even if Higgs boson momentum
 lies
 in these cones it is possible that momentums of the Higgs boson decay
 products (or part of them) will be placed beyond these cones and
 if we do not
 exclude such events (i.e. heavy fermions from Higgs boson
 decays with missing energy from undetected photons) we
 mustn't in
 principle cut these cones.

For the total cross sections where we take into account  mass of
$\mu$-meson we obtain  the following result
\footnote{Despite $ \sqrt{s},m_H \gg m_{\mu} $ we do not neglect
the terms
proportonal $m^2_{\mu}$ because the terms
$(t-m^2_{\mu})^{-2}$,$(u-m^2_{\mu})^{-2}$, in
formulas (5),(8) contain  after integration over $\cos \theta$
singularity $m^{-2}_{\mu}$.Terms of the higher degrees of
 $m_{\mu}$ in numerators may be neglected.
The author expresses his sincere gratitude to
F.Cuypers for this note.}:

\begin{equation}
\label{A7}
 \sigma(\mu^+ \mu^-\rightarrow H^0\gamma)=\frac{\pi \alpha^2}{2sin^2
 \theta_W} \frac{m^2_{\mu}}{m_W^2}F_{S,P}^2 \frac{1}{s-m_H^2}((1+
 \frac{m_H^4}{s^2}) (\log( \frac{s}{m^2_{\mu}})-2 \frac{m_H^2}{s}).
\end{equation}

It must be noted that our calculations  are valid only at
$\sqrt{s}-m_H \gg \Gamma_H$ where $\Gamma_H$ is the total width
of the $H^0$-bosons.

In  general, near threshold
it  is  necessary  to consider virtual
 $H^0$-bosons with finite width in propagator and
 also take into account radiative corrections
which cancell infrared singularity in the process (3).

Analogous situation takes place also for $Z^0$-boson
production (see e.g. ref. \cite{NT} and
references therein)
in the vicinity of  peak where it is  also necessary  to consider the finite
width of $Z^0$-boson and take into account QED-correction which
cancell  infrared singularities of the process
$e^+e^-\rightarrow Z^{0 {\star}}\gamma$.

Besides the contribution to the processes (2),(3) from the Fig.1
there is also a contribution from loops \cite{BPR,ABD} with virtual $W^{\pm}$-and
$t$-quarks and with other heavy particles in various extensions
of the Standard Model such as contributions from squarks, charged Higgs
bosons, chargino \cite{WY}.

It is of interest to compare both tree and loop contributions.For
example, within the Standard Model
essentially below $m_Z$ as seen of the Fig.2  and from Fig.8 of
Ref.\cite{BPR} the tree contribution  exceeds the loop
contribution.At energies
 above $m_Z$ loop contribution dominate over the tree. For
 instance, at $\sqrt{s}=200 GeV \gg m_H $
  as seen from Fig.2 and from figures of Ref.\cite{BPR}
 standard loop contribution exceeds the tree contribution .

On the other hand, at $\sqrt{s} \gg 2m_t$ loop contribution
also decreases  faster than tree contribution, because loop
integrals contain additional
 degree of $s^{-1}$ at $\sqrt{s} \gg 2m_t$.

 Analogous situation
 takes place within the MSSM, at $\sqrt{s} <m_Z$ the
 tree contribution exceeds the loop contribution.

In the MSSM loop contribution to the process (2) depends on
many unknown parameters such as SUSY particle
masses, $\tan\beta$,various SUSY mass parameters containing in
the couplings of Higgs and $Z^0$-bosons with SUSY particles and
the cross section of the loop contribution to the process (2)
 within the MSSM may be increased or
decreased up to two orders in comparision with the loop
contribution
in the Standard model (see Fig.3-11 in the ref. \cite{WY}),
whereas the tree contribution is enhanced by factors
$(\tan\beta)^2$,$(\frac{\cos\alpha}{\cos\beta})^2$,$(\frac{\sin\alpha}{\cos\beta})^2$,(see
Appendix A),where $\tan\beta$ may be as high as $\sim
\frac{m_t}{m_b}\approx 35$ \cite{OP},\cite{GR}
(which enhanced the number of events $H^0_{i} \gamma $ by
1225 times in comparision with  result of Fig.2 ).As seen from
 Fig.3-11 of the ref. \cite{WY}  at $\sqrt{s}=200 GeV $  at the
 wide range of SUSY paprameters considered in \cite{WY}the ratio
$ \sigma(e^+ e^-\rightarrow H^0\gamma)/ \sigma(e^+
 e^-\rightarrow \mu^+\mu^-)/ <10^{-4}$, whereas the same ratio
 for tree contribution at large $\tan\beta$ and far from
 threshold is about
 $\frac{m^2_{\mu}}{m_W^2}\tan^2\beta(\log(
 \frac{s}{m^2_{\mu}})$.For example, at $\tan\beta=20$ this ratio
 for tree contribution is about 0.01 and consequently exceeds loop
 contirbution even at $ \sqrt{s} \gg m_H$.

It must be noted also ,that near threshold (i.e. at $\sqrt{s}
\sim m_H$) tree contribution as seen from formula (7) enhanced as
$\sigma_{tree} \sim 1/(s-m^2_H)$
whereas loop contribution is suppresed as $\sigma_{loop} \sim (s-m^2_H)^3$(see also Fig.8,10 of the Ref.
\cite{WY}).

At $\sqrt{s}$ much larger than masses of particles
( $W^{\pm}$-boson,$t$-quark, charged Higgs boson and their
 superpartners) in loop contributions as well as in the Standard
 Model are supressed by
 additional degree of $s^{-1}$ from loop integrals.

The number of Higgs bosons produced in  the process (2)
 is shown in Fig.2 for case of yearly luminosity
 $L=1000 fb^{-1}$ and at $F=1$
(the muon colliders with luminosities of order  $1000 fb^{-1}$ and
$\sqrt{s}=4 TeV$ has been
considered e.g. in ref. \cite{P}).

              {\bf 3. The $ \mu^+e^-\rightarrow H^0_{fc}\gamma,
              e^+ e^-\rightarrow H_{fc}^0\gamma$ processes}

For the differential cross section for both scalar and
pseudoscalar flavour changing Higgs boson cases we obtain the following result:

\begin{equation}
\label{A8}
\frac{d\sigma( \mu^+ e^-\rightarrow H^0_{fc}\gamma)}{dt}=
\frac{1}{4}\alpha(h^{\mu e}_{S,P})^2\frac{1}{s^2}
(\frac{(m^4_H+s^2)}{(t-m^2_{\mu})(u-m^2_e)}-2m^2_H(\frac{m^2_{\mu}}{(t-m^2_{\mu})^2}+\frac{m^2_e}{(u-m^2_e)^2})),
\end{equation}
where $m_e$ is electron mass.

For the total cross sections we obtain  the following result:

\begin{equation}
\label{A9}
 \sigma(\mu^+ \mu^- \rightarrow H^0_{fc} \gamma)=2\alpha(h^{\mu e}_{S,P})^2
 \frac{1}{s-m_H^2}((1+
 \frac{m_H^4}{s^2}) (\log( \frac{s}{m_{\mu}m_e})-2 \frac{m_H^2}{s}).
\end{equation}

It must be noted, that all the above mentioned remarks concerning collinear
and infrared singularities, missing photons near colliding beam direction and
details of calulation connected with nonzero masses of the
electron and muon remain true also in cases of the processes
(3),(4).
The cross section of the processes $\mu^+\mu^-,
e^+ e^-\rightarrow H_{fc}^0\gamma$ may be obtained by
replacements $m_{\mu}\rightarrow m_e$,$m_{\mu}\rightarrow m_e$
respectively.

It must be noted that although in  (A7) are considered the interaction of
flavour changing pure scalar and pseudoscalars whereas in (A8)
$\tilde{\nu}$ is the  mixing of scalar and pseudoscalar,the cross
sections of the $\tilde{\nu}+\gamma$ and
$H^0_{fc}(P^0_{fc})+\gamma$ production
 at the same Yucawa
couplings (i.e. at $\lambda=h$) are the same
if we neglect terms $O(m^4_{\mu})$  and higher degrees of $m_{\mu}$.

The number of flavour changing Higgs bosons produced in  the
process (3) is shown in Fig.3 at $h^{\mu
 e}_{S,P}=10^{-3}$ .

 As well  as in the case of reaction (2)  the
processes (3),(4) are most effective at low $\sqrt{s} \sim m_H$.

{\bf 4. Aknowledgements}
The author expresses his sincere gratitude to
I.G.Aznauryan, R.P.Grigoryan and K.A.Ispiryan for helpful discussions.

{\bf Note Added}

When this paper has been completed the author received the
information that neutral scalar Higgs boson production with photon
 within the Standard Model
$\mu^+\mu^-\rightarrow H^0\gamma$ has been also studied in
[36],[37].
It must be noted, we disagree with formulas (3),(4) of ref. [36].
The nonlogarithmic terms in our result (formula (7))  and in
formula (4) of the ref.\cite{LT} are different even after
neglecting terms of order $O(m^2_{\mu})$ and higher in the last formula.

\indent
\newpage
\setcounter{equation}{0}
\appendix{{\bf Appendix A}}

\renewcommand{\theequation}{A.\arabic{equation}}
\indent

We parametrized the
model indepedent lagrangian of scalar and pseudoscalar
 Higgs boson ($H^0=S^0,P^0$) interaction with
fermions by the following way:
\begin{equation}
\label{A12}
{\cal L}=  i\frac{gm_f}{2m_W}F_S\bar{f}fS^0+
\frac{gm_f}{2m_W}F_P\bar{f}\gamma_5fP^0
\end{equation}

In the  Standard Model there is only one physical scalar ($F_S=1$)
and pseudoscalars are absent($F_P=0$).

In the MSSM, the Higgs sector
contains two doublets of Higgs bosons with opposite hypercharge
($Y= \pm 1$ ).

After spontaneous symmetry breaking the
following physical states appear:charged Higgs bosons
 $H^\pm$, and three neutral ones,
 $H^0_1, H^0_2, H^0_3 $.

At tree level  the masses of scalars $H^0_1, H^0_2$
 and an angle $\alpha$ (which described the mixing of scalar
 states) are being expressed through the mass of
 pseudoscalar $H^0_3$ and $\ tan\beta=\frac{v_2}{v_1}$ where
 $v_2$,$v_1$  are both doublets vacuum expectations by  following
 relationships:

\begin{equation}
\label{A13}
m^2_{H_1,H_2}=1/2\left[m^2_{H_3}+m^2_Z\pm((m^2_{H_3}+m^2_Z)^2
-4 m^2_Z m^2_{H_3}\cos^2 2\beta)^{1/2}\right]
\end{equation}

\begin{equation}
\label{A14}
\tan2\alpha=\frac{m^2_{H_3}+m^2_Z}{m^2_{H_3}-m^2_Z} \tan2\beta.
\end{equation}

It follows from (A1) that MSSM guarantes the existence of, at
least, one light Higgs boson with
$m_{H_2}<m_{Z}$.

Interactions of the $H^0_i$-bosons with fermions are described by
lagragian:

\begin{equation}
\label{A15}
{\cal L}=
i\frac{gm_d}{2m_W}\frac{\cos\alpha}{\cos\beta}\bar{d}dH_1^0+
i\frac{gm_d}{2m_W}\frac{\sin\alpha}{\cos\beta}\bar{d}dH_2^0+
\frac{gm_d}{2m_W}\tan\beta\bar{d}\gamma_5dH_3^0
\end{equation}

At $\ tan\beta \gg 1$ the mass relation (A2),(A3) and formula
(A4) are strongly reduced:

\begin{equation}
\label{A16}
m_{H_2}=m_{H_3},m_{H_1}=m_Z ,F_{H_2}=\tan \beta \gg F_{H_1}
at\,\, m_{H_3}>m_Z,
\end{equation}

\begin{equation}
\label{A17}
m_{H_2}=m_{H_3} ,m_{H_1}=m_{H_3},F_{H_2}=\tan \beta \gg F_{H_1}
at \,\,  m_{H_3}>m_Z .
\end{equation}
It must be noted, that radiative
corrections \cite{OYY}-[33] can strongly change
relations (A1),(A2) however
in the large $\tan\beta$ limit and at $m_{H_3}<m_Z$  or at
$m_{H_3} \gg m_Z$ formulas (A2),(A3),(A5) hold approximately true
even after taking into account the radiative corrections.

We parametrized the model indepedent lagrangian of flavour
changing scalar and pseudoscalar Higgs bosons ($H^0_{fc}=S^0_{fc},P^0_{fc}$)
interaction with fermions by following way (see e.g. \cite{KRSZ}):
\begin{equation}
\label{A18}
{\cal L}=  ih_S^{ij}\bar{f_i}f_jS^0_{fc}+
ih_P^{ij}\bar{f_i}\gamma_5f_jP^0_{fc}+H.c.
\end{equation}

Analogous interactions may also derive from $R$-parity braking
part of the superpotential in the above mentioned theories with
$R$-parity violation.In four-component Dirac notations $R$-parity
violating interactions has the following form :

\begin{equation}
\label{A19}
{\cal L}=  \lambda_{ijk}(\bar{l^i}\tilde{\nu}_jl^k_R+neutrino
interactions)+H.c.
\end{equation}

where $l^i,\tilde{\nu}^j$ are charged lepton and scalar neutrino
respectively.

sections (8).

\newpage

\begin{figure}
\begin{center}
\epsfxsize=10.cm
\leavevmode\epsfbox{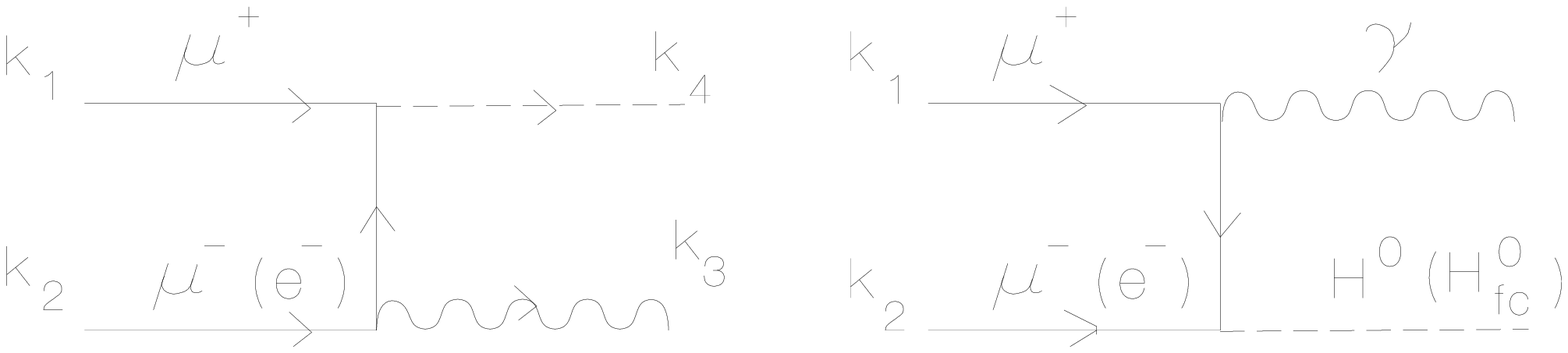}
\end{center}
\caption{Diagrams corresponding to the processes (2),(3).}
\end{figure}

\begin{figure}
\begin{center}
\epsfxsize=10.cm
\leavevmode\epsfbox{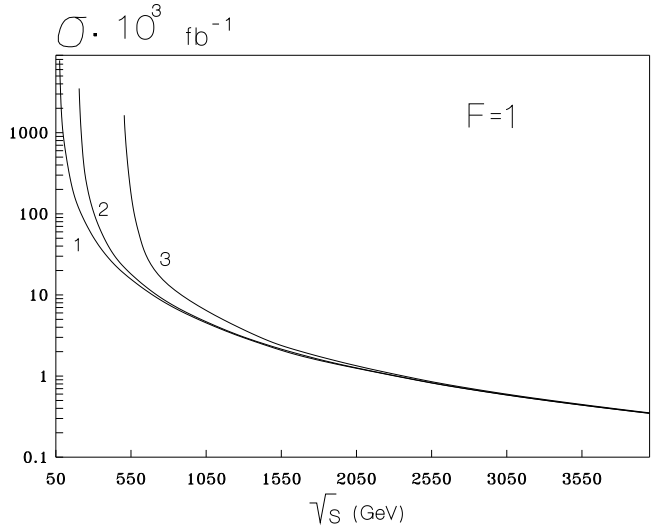}
\end{center}
\caption{Number of Higgs bosons produced in the process (2)
as a function of $  square$ $ root$ $s$ at fixed $m_H$.
Curves 1,2,3 correspond to the $m_H=70,200,500 GeV $ respectively.}
\end{figure}

\begin{figure}
\begin{center}
\epsfxsize=10.cm
\leavevmode\epsfbox{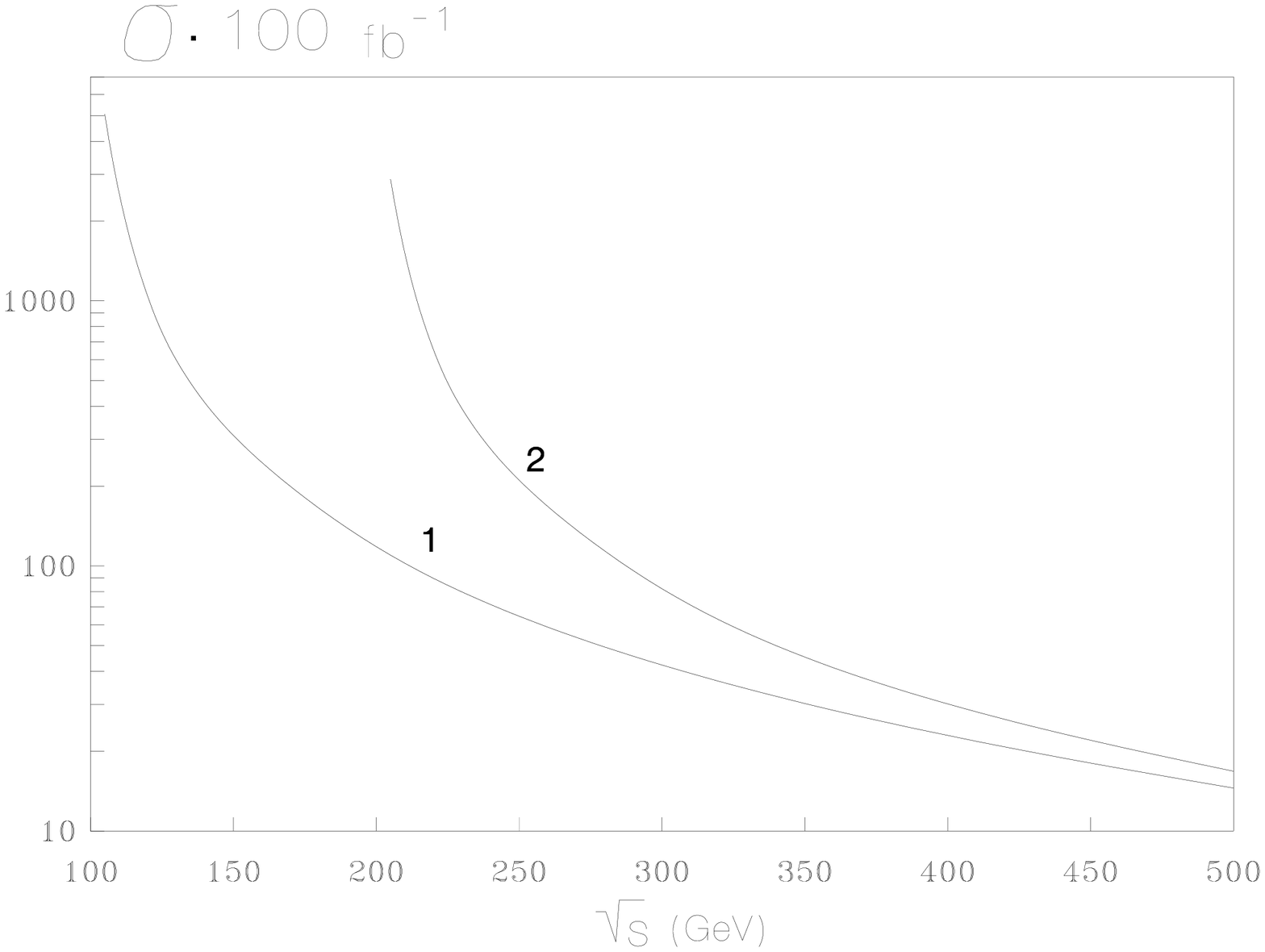}
\end{center}
\caption{ Number of Higgs bosons produced in the process (3)
as a function of $square$ $ root$ $s$ at fixed $m_H$ at $h^{\mu e}_{S,P}=10^{-3}$.
Curves 1,2 correspond to the $m_H=100,200 GeV $ respectively.}
\end{figure}

\begin{figure}
\begin{center}
\epsfxsize=10.cm
\leavevmode\epsfbox{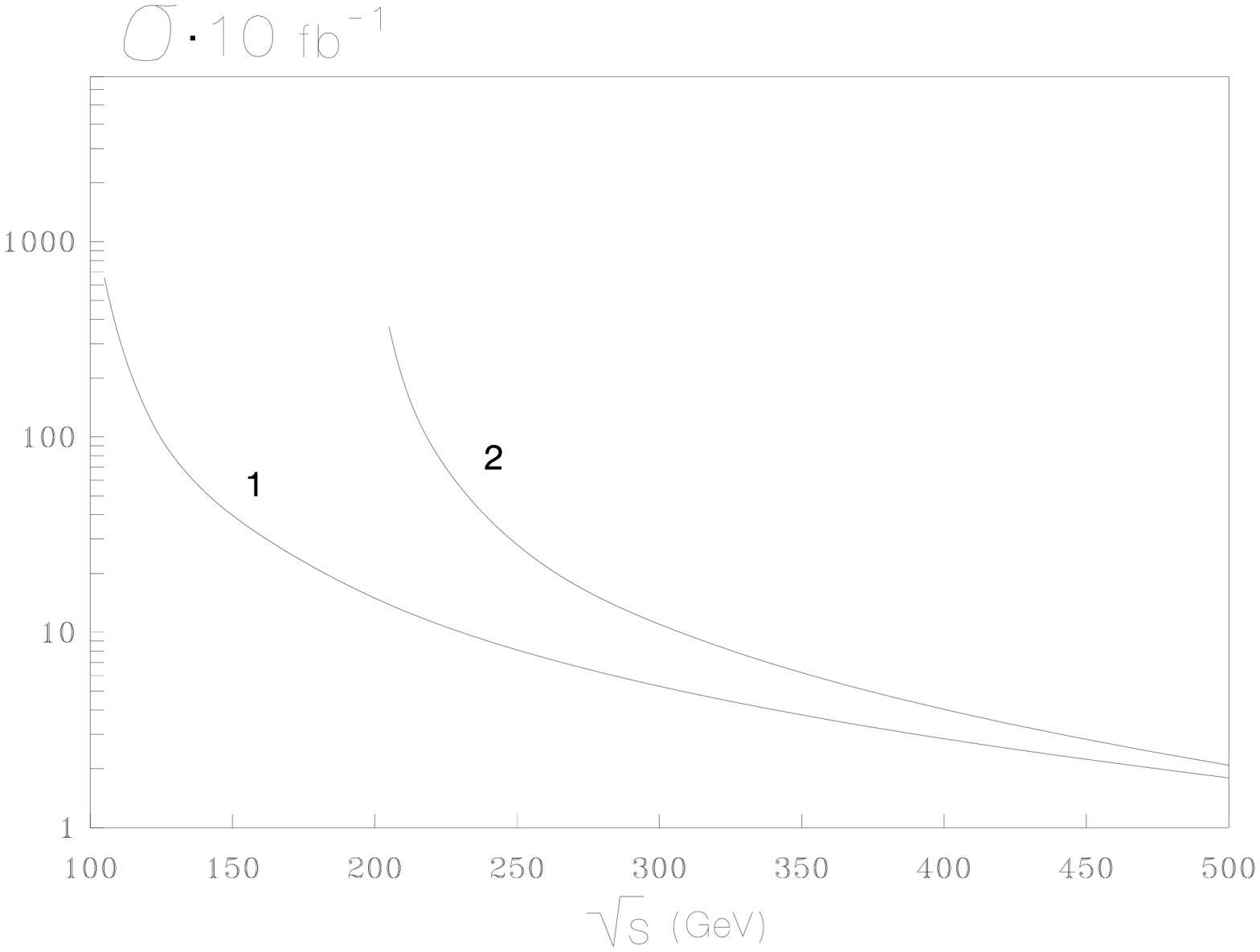}
\end{center}
\caption{ Number of Higgs bosons produced in the process (3)
as a function of $square$ $root$ $s$ at fixed $m_H$ at $h^{\mu e}_{S,P}=10^{-3}$.
Curves 1,2 correspond to the $m_H=100,200 GeV $ respectively.}
\end{figure}
\end{document}